\documentclass[pre,twocolumn,twoside,byrevtex,superscriptaddress,floatfix]{revtex4-1}

\usepackage[realmainfile]{currfile}
\makeatletter
\let\ftype@table\ftype@figure
\makeatother

\usepackage{color}

\usepackage{graphicx,epsfig,verbatim,enumerate}
\graphicspath{{figures/local/}{figures/local/emojis/}{figures/local/tables/}{figures/local/timeseries/}}

\usepackage{amssymb,amsmath}
\usepackage{ifthen}

\usepackage{longtable}

\usepackage{mathtools}

\newboolean{twocolswitch}

\newcommand{\sindex}[1]{}
\newcommand{\nindex}[1]{}

\newcommand{\www}[1]{\url{#1}}

\usepackage{lettrine}

\setboolean{twocolswitch}{true}

\begin{document}

\title{\protect
  How the world's collective attention is being paid to a pandemic:\\
  COVID-19 related $n$-gram time series for 24 languages on Twitter
}

\author{
  \firstname{Thayer}
  \surname{Alshaabi}
}

\thanks{The first three authors contributed as a team and are listed in alphabetical order.}

\affiliation{
  Computational Story Lab,
  Vermont Complex Systems Center,
  MassMutual Center of Excellence for Complex Systems and Data Science,
  Vermont Advanced Computing Core,
  University of Vermont,
  Burlington, VT 05401.
  }

\author{
  \firstname{Michael V.}
  \surname{Arnold}
}

\thanks{The first three authors contributed as a team and are listed in alphabetical order.}

\affiliation{
  Computational Story Lab,
  Vermont Complex Systems Center,
  MassMutual Center of Excellence for Complex Systems and Data Science,
  Vermont Advanced Computing Core,
  University of Vermont,
  Burlington, VT 05401.
  }

\author{
  \firstname{Joshua R.}
  \surname{Minot}
}

\thanks{The first three authors contributed as a team and are listed in alphabetical order.}

\affiliation{
  Computational Story Lab,
  Vermont Complex Systems Center,
  MassMutual Center of Excellence for Complex Systems and Data Science,
  Vermont Advanced Computing Core,
  University of Vermont,
  Burlington, VT 05401.
  }

\author{
  \firstname{Jane Lydia}
  \surname{Adams}
}

\affiliation{
  Computational Story Lab,
  Vermont Complex Systems Center,
  MassMutual Center of Excellence for Complex Systems and Data Science,
  Vermont Advanced Computing Core,
  University of Vermont,
  Burlington, VT 05401.
}

\author{
  \firstname{David Rushing}
  \surname{Dewhurst}
}

\affiliation{
  Computational Story Lab,
  Vermont Complex Systems Center,
  MassMutual Center of Excellence for Complex Systems and Data Science,
  Vermont Advanced Computing Core,
  University of Vermont,
  Burlington, VT 05401.
  }

\author{
   \firstname{Andrew J.}
   \surname{Reagan}
 }
 
 \affiliation{
   MassMutual Data Science,
   Amherst,
   MA 01002.
   }

\affiliation{
  Computational Story Lab,
  Vermont Complex Systems Center,
  MassMutual Center of Excellence for Complex Systems and Data Science,
  Vermont Advanced Computing Core,
  University of Vermont,
  Burlington, VT 05401.
  }

\author{
  \firstname{Roby}
  \surname{Muhamad}
  }

\affiliation{
International Class Program,
Faculty of Social and Political Sciences,
University of Indonesia,
Jakarta, Indonesia. 
}

\author{
  \firstname{Christopher M.}
  \surname{Danforth}
}

\affiliation{
  Computational Story Lab,
  Vermont Complex Systems Center,
  MassMutual Center of Excellence for Complex Systems and Data Science,
  Vermont Advanced Computing Core,
  University of Vermont,
  Burlington, VT 05401.
  }

\affiliation{
  Department of Mathematics \& Statistics,
  University of Vermont,
  Burlington, VT 05401.
  }

\author{
  \firstname{Peter Sheridan}
  \surname{Dodds}
}

\email{Corresponding author: peter.dodds@uvm.edu}

\affiliation{
  Computational Story Lab,
  Vermont Complex Systems Center,
  MassMutual Center of Excellence for Complex Systems and Data Science,
  Vermont Advanced Computing Core,
  University of Vermont,
  Burlington, VT 05401.
  }

\affiliation{
  Department of Mathematics \& Statistics,
  University of Vermont,
  Burlington, VT 05401.
  }

\date{\today}

\begin{abstract}
  \protect
  In confronting the global spread of the coronavirus disease COVID-19 pandemic
we must have coordinated medical, operational, and political responses.
In all efforts, data is crucial.
Fundamentally, and in the possible absence of a vaccine for 12 to 18 months,
we need universal, well-documented testing for both the presence of the disease as
well as confirmed recovery through serological tests for antibodies,
and we need to track major socioeconomic indices.
But we also need auxiliary data of all kinds, including data related
to how populations are talking about the unfolding pandemic through news and stories.
To in part help on the social media side, we curate a set of 1000 day-scale time series of
1-grams across 24 languages on Twitter that are most `important'
for March 2020 with respect to March 2019.
We determine importance through our allotaxonometric instrument, rank-turbulence divergence.
We make some basic observations about some of the time series, including a comparison to
numbers of confirmed deaths due to COVID-19 over time.
We broadly observe across all languages a peak for
the language-specific word for `virus' in January followed
by a decline through February and a recent surge through March.
\textbf{The world's collective attention dropped away while the virus spread out from China.}
We host the time series on Gitlab, updating them on a daily basis while relevant.
Our main intent is for other researchers to use these time series to enhance whatever
analyses that may be of use during the pandemic as well as for retrospective investigations.
 
\end{abstract}

\pacs{89.65.-s,89.75.Da,89.75.Fb,89.75.-k}

\maketitle

\nocite{li2020a,kraemer2020a}
\nocite{dong2020a,xu2020a}
\nocite{cinelli2020a,chen2020a,lampos2020a}


\section{Overview}
\label{sec:covid19.overview}

Understanding how major disasters affect the wellbeing of populations both
in real time and historically is of paramount importance.
We especially need real-time measurement to enable policy makers in health systems and government
to gauge the immediate situation and evaluate scenarios,
and for researchers to model possible future trajectories of social systems.

In this short piece,
we describe how we select languages and 1-grams relevant to the
time period of the present COVID-19 pandemic;
show example time series plots for `virus', including a visual comparison with
COVID-19 confirmed case and death numbers;
and
describe the data sets, figures, and visualizations
for 24 languages that we share online.

\section{Selection of languages and $n$-grams}
\label{sec:covid19.methods}

\begin{table}[tp!]
  \begin{tabular}{rlc}
    Rank & Language & Code \\
    \hline
    1 & English & en \\
    2 & Spanish & es \\
    3 & Portuguese & pt \\
    4 & Arabic & ar \\
    5 & Korean & ko \\
    6 & French & fr \\
    7 & Indonesian & id \\
    8 & Turkish & tr \\
    9 & German & de \\
    10 & Italian & it \\
    11 & Russian & ru \\
    12 & Tagalog & tl \\
  \end{tabular}
  \qquad
  \begin{tabular}{rlc}
    Rank & Language & Code \\
    \hline
    13 & Hindi & hi \\
    14 & Persian & fa \\
    15 & Urdu & ur \\
    16 & Polish & pl \\
    17 & Catalan & ca \\
    18 & Dutch & nl \\
    19 & Tamil & ta \\
    20 & Greek & el \\
    21 & Swedish & sv  \\
    22 & Serbian & sr \\
    23 & Finnish & fi \\
    24 & Ukrainian & uk \\
  \end{tabular}
  \caption{
    The 24 languages for which we provide COVID-19 related Twitter time series.
    See Tab.~\ref{tab:covid19.1gramcounts} for total counts of 1-grams per language
    in our data set from 2019/09/01 through to 2020/03/23.
    }
  \label{tab:covid19.languages}
\end{table}

Our primary aim here is to generate a particular data stream that may be of
help to other researchers:
A principled set of 1-gram time series across major languages
used on Twitter and news-relevant for March, 2020.
Our work is complementary to extant efforts to enable research
on the COVID-19 pandemic~\cite{li2020a,kraemer2020a}
by gathering and sharing epidemiological data~\cite{dong2020a,xu2020a},
economic data, and internet and social media data~\cite{cinelli2020a,chen2020a,lampos2020a}.

We base our curation on our work in two of our previous papers~\cite{alshaabi2020a,dodds2020a},
and we draw from a database of approximately 10\% of all tweets from 2008/09/09 to present.

Our process of obtaining salient 1-grams for March 2020 comprises two steps.

First, in~\cite{alshaabi2020a},
we used the Language Identification (LID) model FastText~\cite{joulin2016a,joulin2016b}
to evaluate all tweets in our historical archive,
finding over 150 languages.
We subsequently extracted GMT day-scale 
Zipf distributions
for 1-, 2-, and 3-grams along with
day-scale $n$-gram time series.
We focus here on 1-grams.
We preserve case where applicable, do not apply any stemming,
and include hashtags, handles, and emojis.

Besides analyzing all tweets (AT), we also separately process what we call
organic tweets (OT): All Twitter messages which are original.
Organic tweets exclude retweets
while including all added text for quote tweets.
In doing so, we are able to carry through a measure of spreadability
for all $n$-grams.
The key threshold we use for spreading is the naive one from biological and social contagion models:
When a 1-gram appears in more retweeted than organic material, we view it as being socially amplified.

Here, we take 24 of the most commonly used languages
on Twitter in 2019, with the provision that we are able to parse them into 1-grams.
For the time being, we are unable to reliably parse Japanese, Thai, and Chinese,
the 2nd, 6th, and 13th most common languages.
We exclude all tweets not assigned a language with sufficient confidence
(an effective 4th ranked collection)
and choose to include Ukranian (29th) over Cebuano (28th).
We list the 24 languages by overall usage frequency in Tab.~\ref{tab:covid19.languages}.

Second, we compare usage of 1-grams in March of 2020 with March 2019
to determine which 1-grams have become most elevated in relative usage.
We do so by using rank-turbulence divergence~\cite{dodds2020a}, an instrument
for comparing any pair of heavy-tailed size distributions of categorical data.
Other well-considered divergences will produce similar lists.

For each language, we take Zipf distributions for each of the first 21 days of March 2020,
and compare them with the Zipf distributions of 52 weeks earlier.
We use rank-turbulence divergence with the parameter $\alpha$ set to 1/3 as this provides
a reasonable fit to the lexical turbulence we observe~\cite{pechenick2017a,dodds2020a}.

\begin{figure*}[tp!]
  \includegraphics[width=1.1\textwidth]{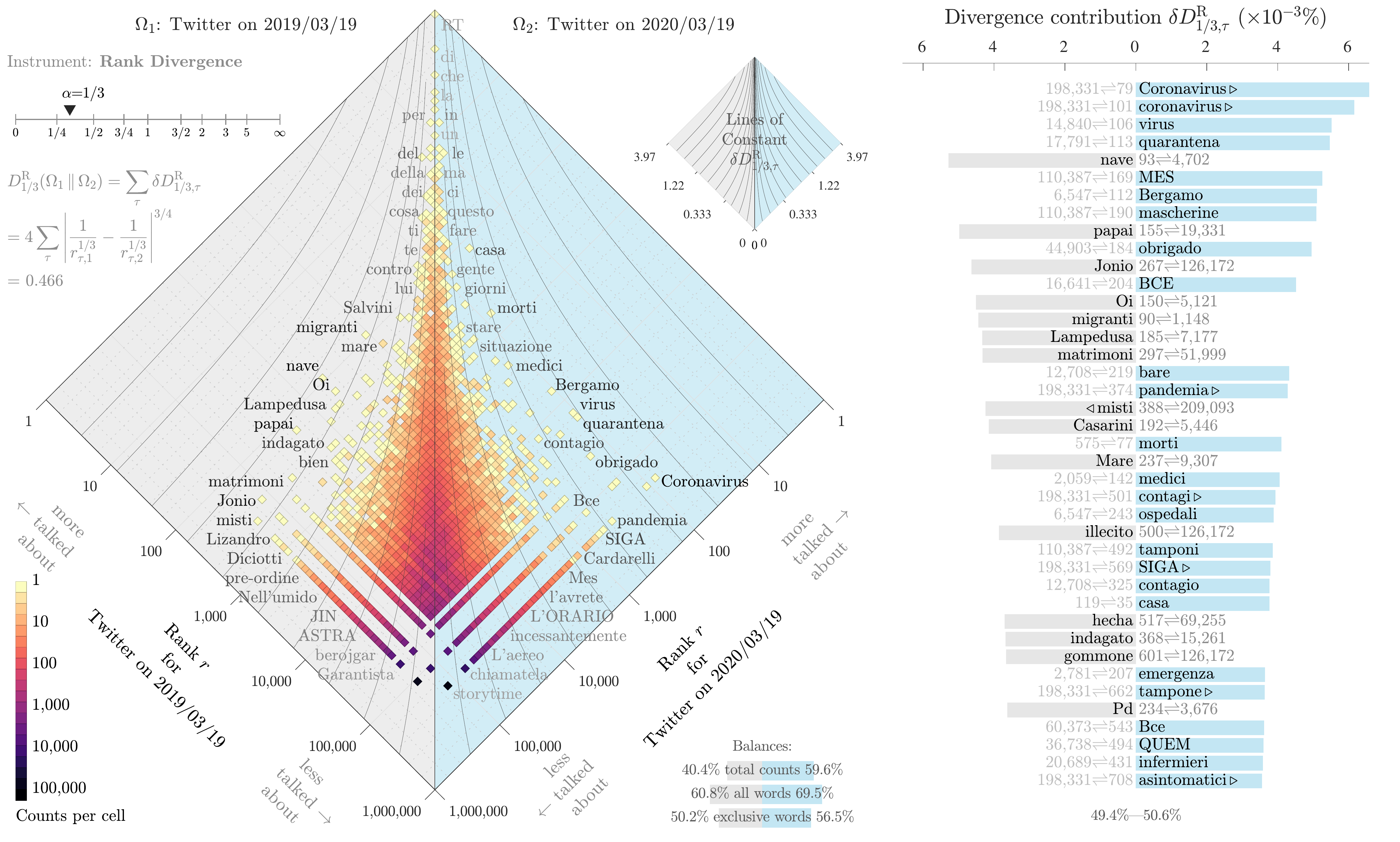}
  \caption{
    Allotaxonograph using rank-turbulence divergence for
    Italian word usage on March 19, 2019 versus March 19, 2020.
    For this visualization, we consider the subset of 1-grams that are formed from latin characters.
    The right hand sides of the rank-rank histogram and the rank-turbulence contribution list
    are dominated by COVID-19 related terms.
    See~\cite{dodds2020a} for a full explanation of our allotaxonometric instrument.
  }
  \label{fig:covid19.allotaxonometer9000-2019-03-19-2020-03-19-rank-div-alpha-third-it}
\end{figure*}

\begin{table*}[tp!]
  \includegraphics[width=.95\textwidth]{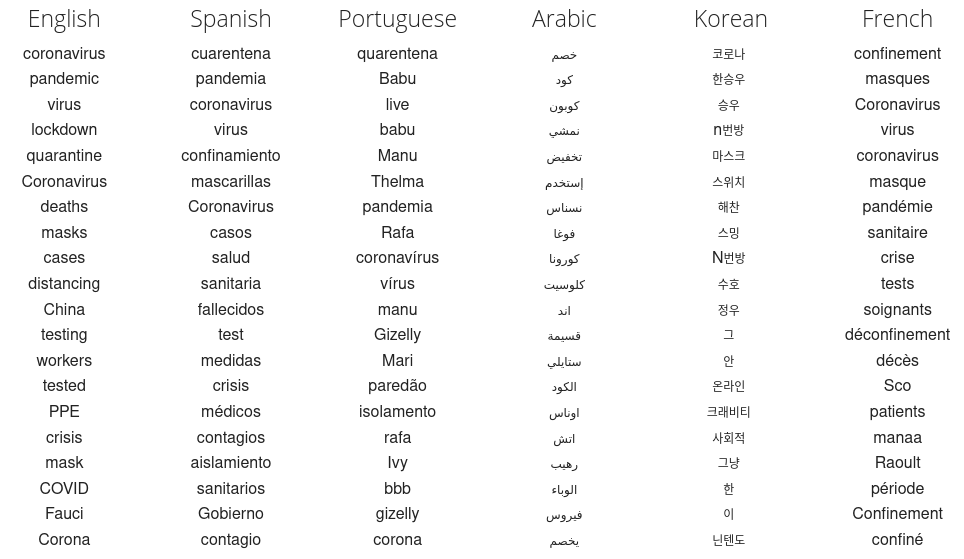}\\
  \includegraphics[width=.95\textwidth]{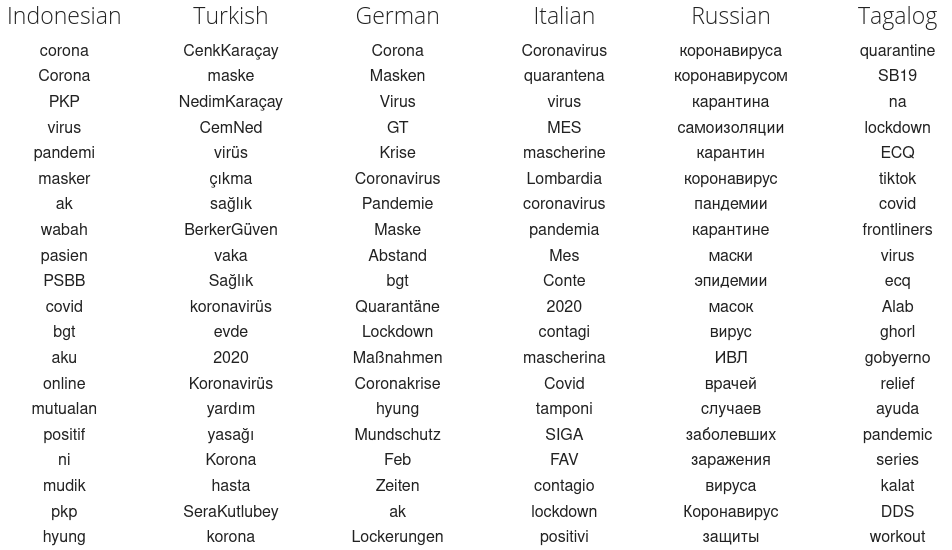}
  \caption{
    Top 20 (of 1,000) 1-grams for our top 12 languages
    for the first three weeks of March 2020 relative to a year earlier.
    Our intent is to capture 1-grams that are topically and culturally important
    during the COVID-19 pandemic.
    While overall, we see pandemic-related words dominate the lists across languages,
    we also find considerable specific variation.
    Words for virus, quarantine, protective equipment, and testing show different orderings
    (note that we do not employ stemming).
    Unrelated 1-grams but important to the time of March 2020 are in evidence; the balance
    of these are important for our understanding of how much the pandemic is being talked about.
    Some 1-grams such as punctuation represent functional changes in the use of Twitter across languages;
    we include them nonetheless.
    To generate these lists we use the allotaxonometric method of
    rank-turbulence divergence to find the most distinguishing 1-grams
    (see Sec.~\ref{sec:covid19.methods},
    Fig.~\ref{fig:covid19.allotaxonometer9000-2019-03-19-2020-03-19-rank-div-alpha-third-it},
    and Ref.~\cite{dodds2020a}).
    We note that these tables are images and cannot be copied;
    see the Sec.~\ref{sec:covid19.data} for data download sites.
  }
  \label{fig:covid19.top1grams-1}
\end{table*}

\begin{table*}[tp!]
  \includegraphics[width=.95\textwidth]{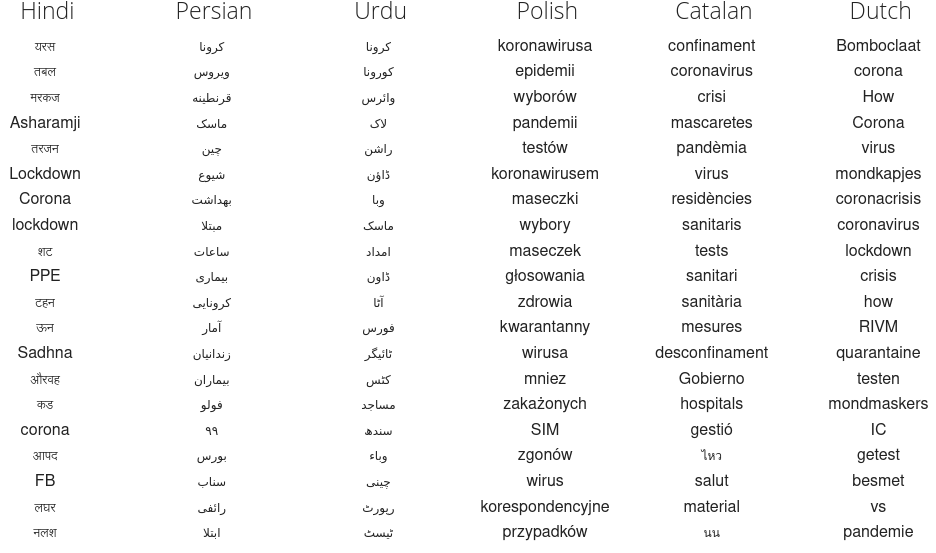}\\
  \includegraphics[width=.95\textwidth]{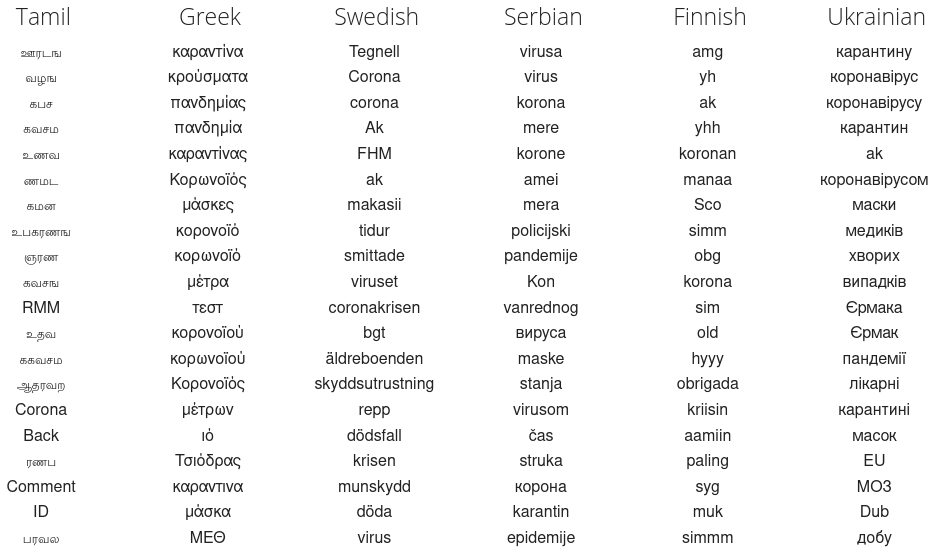}
  \caption{
    Continuing on from Fig.~\ref{fig:covid19.top1grams-1}:
    Top 20 1-grams for the second 12 of 24 languages we study
    for March 2020 relative to March 2019.
  }
  \label{fig:covid19.top1grams-2}
\end{table*}

\begin{figure*}[tp!]
    \includegraphics[width=\textwidth]{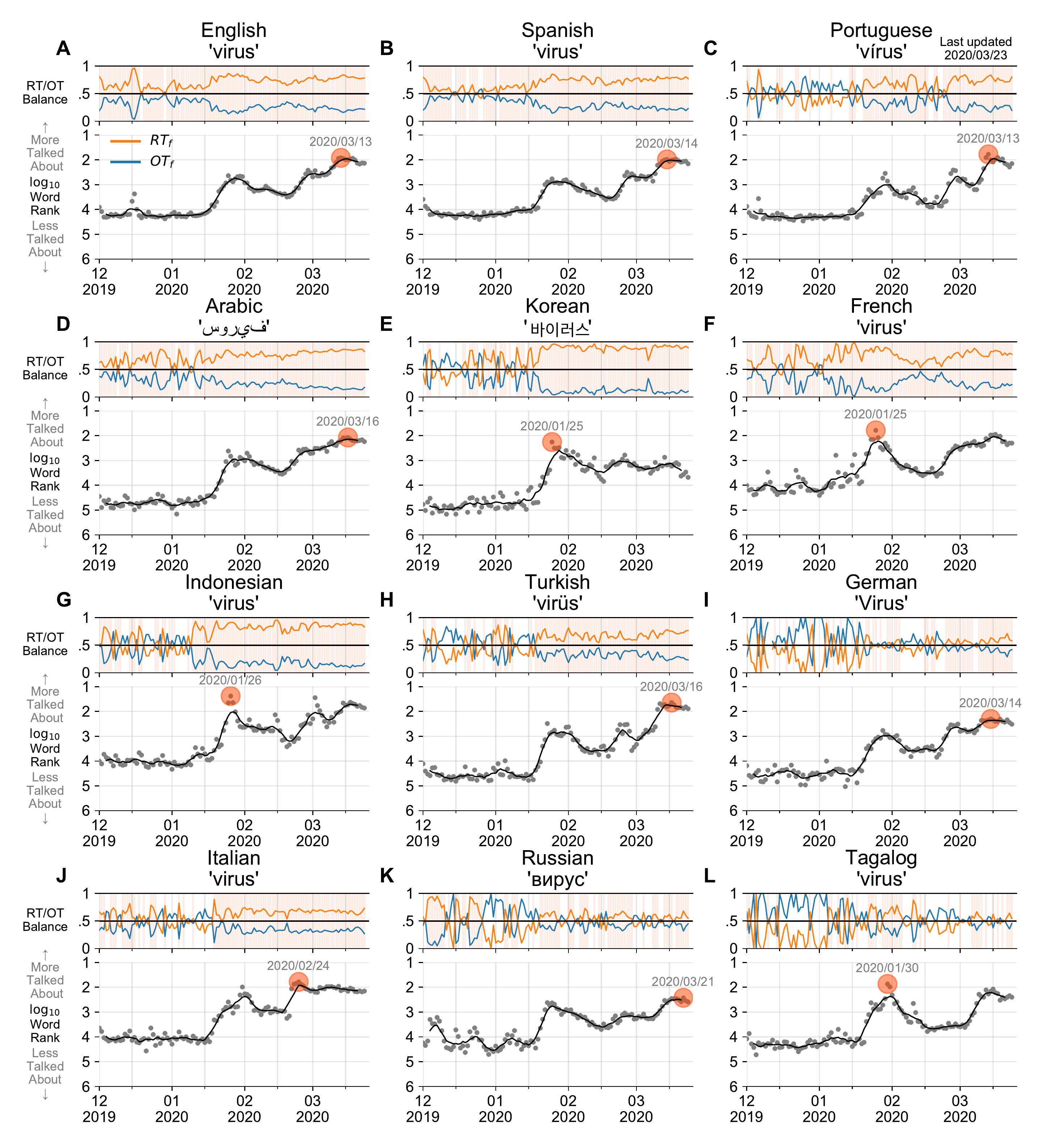}
    \caption{
      Usage rank time series for the word
      `virus' in the top 12 of the 24 languages we study here.
      \textbf{The major observation is that the world's attention peaked early in late January around the news
      of an outbreak of a new infectious disease in Wuhan,
      declining through well into February before waking back up.}
      The main plots in each panel show usage ranks at the day scale (GMT)
      (we use common sense ranking meaning the most frequently used 1-gram has rank $r=1$).
      The solid lines indicating smoothing with a one week average (centered).
      The plots along the top of each panel show the relative fractions of each 1-gram's daily counts
      indicating as to whether they appear in retweets (RT, spreading) or organic tweets (OT, new material).
      The background shading shows when the balance favors spreading---story contagion.
      See Fig.~\ref{fig:covid19.virus2_ranks} for the next 12 languages, as well as
      Sec.~\ref{sec:covid19.timeseries} for general discussion.
      See also the companion plots of relative frequency time series
      in Fig.~\ref{fig:covid19.virus1}.
    }
    \label{fig:covid19.virus1_ranks}
\end{figure*}

\begin{figure*}[tp!]
    \includegraphics[width=\textwidth]{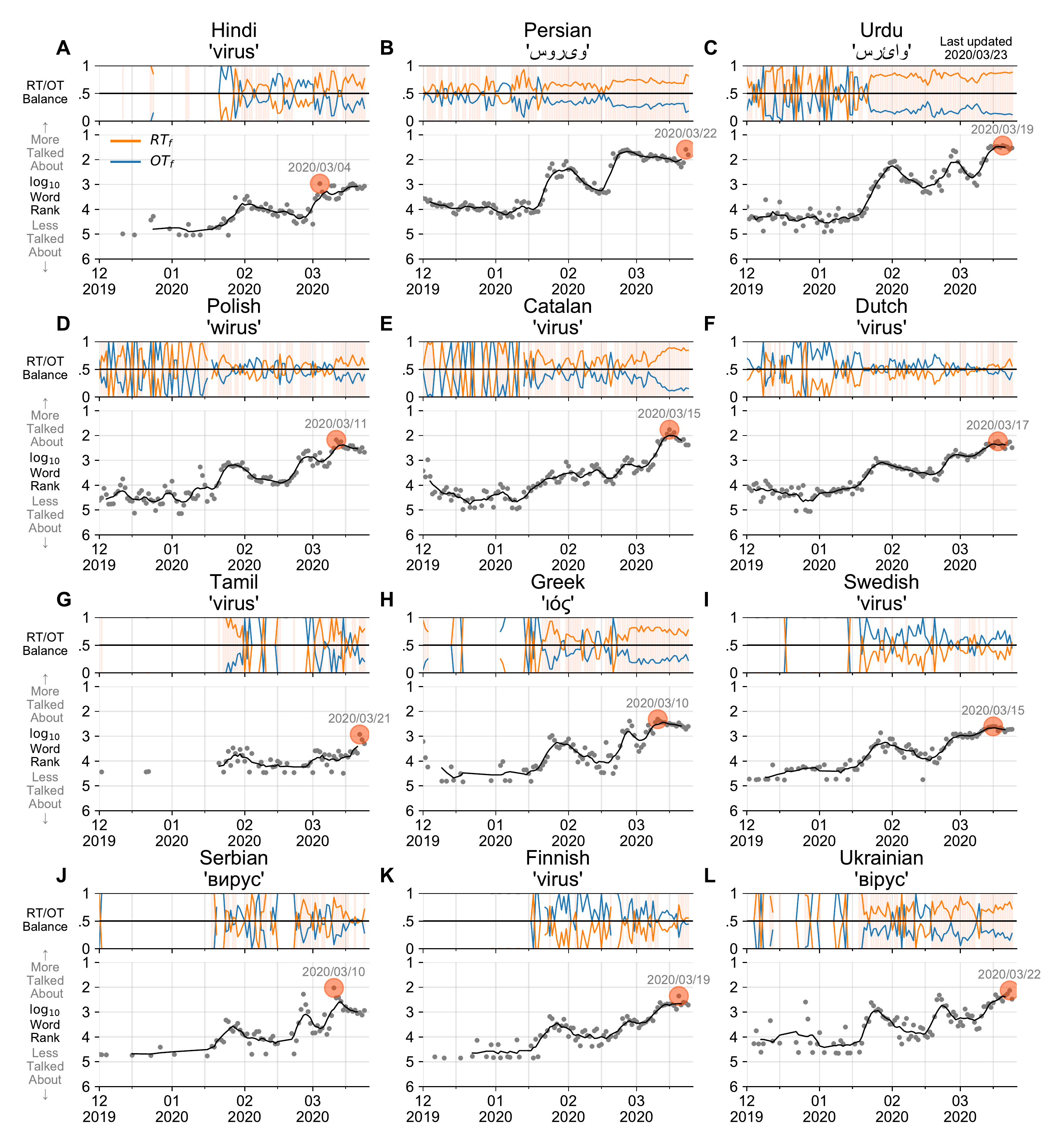}
    \caption{
      Following on from Fig.~\ref{fig:covid19.virus1_ranks},
      usage rank time series for the word
      `virus' in the second 12 of the 24 languages.
      See also the companion plots of relative frequency time series
      in Fig.~\ref{fig:covid19.virus2}.
    }
    \label{fig:covid19.virus2_ranks}
\end{figure*}

\begin{figure*}[tp!]
    \includegraphics[width=\textwidth]{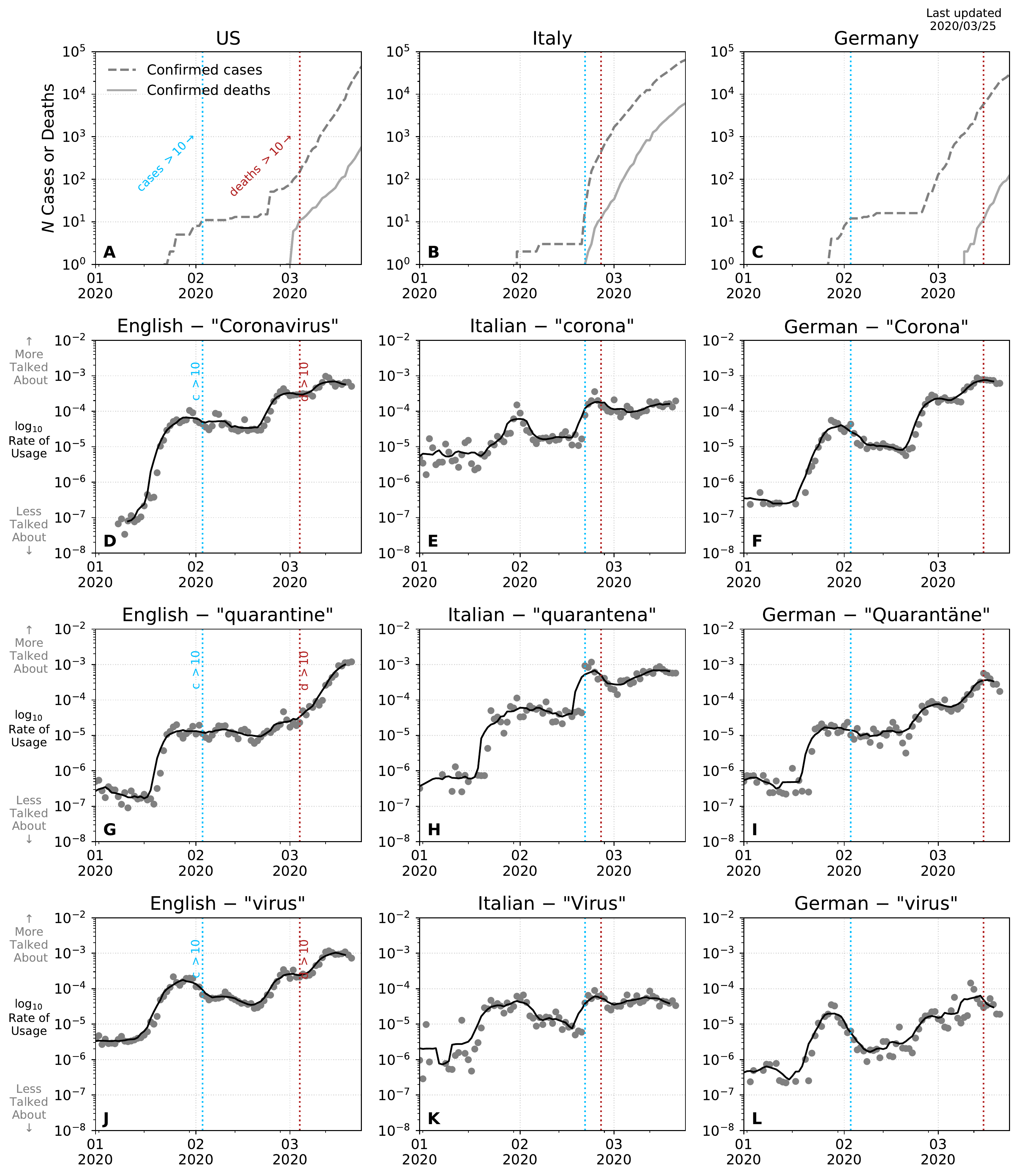}
    \caption{
      Time series for known case loads and death counts for
      the United States, Italy, and Germany,
      compared with 1-grams
      for `Coronavirus', `quarantine', and `virus'
      in English, Italian, and German.
      We note that the counts
      in plots A, B, and C are underestimates, more so for cases than
      deaths, and errors are unknown.  We sourced data for confirmed
      cases and fatalities from Johns Hopkins University Center for
      Systems Science and Engineering's COVID-19
      project~\cite{dong2020a}.
      Starting on 2020/01/22,
      the project's data has been collected from national and regional health
      authorities across the world. The data is augmented by case
      reports from medical associations and social media posts---these
      later sources are validated against official records before
      publication.
      For the present piece, we use daily summary files for case
      counts and fatalities, although an API and online dashboard are
      available for more up-to-date reports.
    }
    \label{fig:covid19.corona_grams_US_Italy_Germany}
\end{figure*}

For an example, we show in
Fig.~\ref{fig:covid19.allotaxonometer9000-2019-03-19-2020-03-19-rank-div-alpha-third-it}
an allotaxonograph for Italian comparing 2019/03/19 and 2020/03/19.
For ease of plotting, we have further chosen to compare the subset of words containing latin characters only.
Words associated with the pandemic dominate the contributions from 2020/03/19.
On the right side of the allotaxonograph,
we see
`Coronavirus', 
`virus',
`Bergamo',
`pandemia',
and
`morti'.

We next combine rank-turbulence divergence contributions for all 1-grams across
the first 21 days of March, and rank 1-grams in descending order.

Finally, for each language, and for each of the top 
1-grams we have identified, we extract
three day-scale time series starting on 2019/09/01:
daily (GMT) counts, ranks, and normalized frequencies.

We repeat all of the above steps for 1-grams derived from organic tweets (OT).

We show the resulting top 20 March-2020-specific 1-grams for the 24 languages
in
Tabs.~\ref{fig:covid19.top1grams-1}
and~\ref{fig:covid19.top1grams-2}.
Overall, we see that the lists are dominated by language specific words 
for
coronavirus
virus,
quarantine,
pandemic,
testing,
and spreading.
Not every language is pandemic focused however with Hindi, Tamil, and Finnish
being three examples.

The microbe emoji, \includegraphics[height=1em]{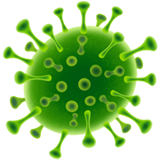},
makes the top 20 in only Spanish and German.
By contrast, and according to the measurements we have used here,
the worried face emoji,
\includegraphics[height=1em]{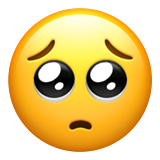},
has become important across many languages in March 2020 relative to March 2019.
It would be natural to see this emoji as being pandemic-related
but in fact, we see from time series that the worried emoji has slowly being increasing
in usage over time for several years
(determining the reasons for which we will leave for a separate line of inquiry).

We emphasize that with our approach, we do not explicitly
determine whether or not a 1-gram is relevant
to COVID-19.
While the pandemic is the global story of March 2020,
there have of course been other major events and
moments in popular culture around the world.
For example, for the United States, the ongoing democratic primary
leads to the 1-gram in English Twitter of `Biden' being prominent.

We also observe variations in punctuation and grammatical structures.
All of these non-pandemic-related elements may of course be filtered out
for individual languages by hand.
While we do not have the capability and bandwidth across our team to perform such
specific tasks,
the degree to which the pandemic is being discussed on Twitter is
of great interest in itself, and our data set will allow for such
examination.

Further, most languages have a strong degree of geographic specificity
(e.g., Finnish for Finland, Portuguese for Brazil),
and we have not
filtered for precise geo-location.
English, Spanish, Arabic, and French are some of the more
geographically distributed languages.

\section{Example time series}
\label{sec:covid19.timeseries}

We provide and
briefly consider two sets of
sample time series based on our data set.

Across Figs.~\ref{fig:covid19.virus1_ranks} and~\ref{fig:covid19.virus2_ranks},
we plot usage rank time series for the word `virus' translated as appropriate
in to each of the 24 languages.
(See Figs.~\ref{fig:covid19.virus1} and~\ref{fig:covid19.virus2}
for normalized rates of usage versions.)

For each language, we show the daily (Zipfian) rank for `virus' in the main panel of each plot.
The pale disk highlights the date of maximum observed rate.
In the secondary time series at the top of each panel,
we show the relative fraction
of 1-gram contained in all tweets (RT) versus organic tweets (OT).
When the RT/OT balance exceeds 50\%, we shade the background to indicate
that the 1-gram is being spread (e.g., retweeted) more than freshly tweeted.

\textbf{
  Alone, the highest ranks for `virus' show the enormity of the pandemic.
  While a common enough word in normal times, `virus' has reached into
  the top 100 ranks across many languages, a region that we have elsewhere
  referred to as the realm of lexical ultrafame~\cite{dodds2019a}.}
Normally only the most basic function words of a language will
populate the top 100 ranks.

In the last few months,
we have seen `virus' rise as high
as $r$=24 in Indonesian (2020/01/26),
$r$=27 in Polish (2020/03/11),
$r$=29 in Urdu (2020/03/22),
$r$=44 in German (2020/03/14),
and
$r$=83 in English (2020/03/13).

In terms of the shapes of the time series for `virus',
most languages show a late January peak consistent with the news from China
of a novel coronavirus disease spreading in Wuhan.
The subsequent drop in usage rate across most of the 24 languages reflects a
global decline in attention being paid to the outbreak.

The Italian time series for `virus' in
Fig.~\ref{fig:covid19.virus1_ranks}J shows an abrupt jump about
three quarters of the way through February,
strikingly just after a drop in RT/OT balance.
Persian has a similar shock jump just after
midway of February 
(Fig.~\ref{fig:covid19.virus1_ranks}B).

A region currently experiencing a major outbreak is Catalonia.
We see in Fig~\ref{fig:covid19.virus2_ranks}E
that `virus' in Catalan shows no early January peak
like most of the other 23 languages, suggesting that even the initial news from China
did not have great impact.

One of the major problems we face with the COVID-19 pandemic
is the unevenness of testing across the world.
South Korea and Iceland have tested early and extensively
while the United States's testing has been uncoordinated and slow to expand.

Urdu's heightened time series for `virus' (Fig~\ref{fig:covid19.virus2_ranks}C)
would seem especially concerning given low numbers coming out of Pakistan
which, as of 2020/03/24, had reported 1,063 cases and 8 deaths~\cite{dong2020a}.
For Indonesia, where testing has also been limited~\cite{dong2020a}
and with peak attention on Twitter coming in January and early focus
on economic issues and evacuation of nationals from Wuhan, a dip in
the rank of `virus' in the second half of February is also worrying (Fig~\ref{fig:covid19.virus1_ranks}G).

As one very simple example of comparing our Twitter times series
with pandemic-related data, in Fig.~\ref{fig:covid19.corona_grams_US_Italy_Germany},
we present plots of confirmed cases and deaths over time
for the United States, Italy, and Germany, along with
time series for
`Coronavirus',
`quarantine',
and
`virus'
(translated for Italian and German).
We indicate where numbers of known cases and known deaths
reach 10 with vertical dotted lines, and note that these
are surely under-reportings of unknown error.
The points which known case loads and known death tolls reach 10 vary
considerably across the three countries, with Italy rapidly moving
from 10 known cases to 10 known deaths.

\section{Data and Sites}
\label{sec:covid19.data}

We share and maintain all data on Gitlab at:\newline
\href{https://gitlab.com/compstorylab/covid19ngrams}{https://gitlab.com/compstorylab/covid19ngrams}.

For each of the 24 languages,
we provide time series for the top 1000 1-grams for both
all tweets (AT) and organic tweets (OT).
Note that in general, a 1-gram in the AT group may not appear in the OT group,
as we start with only 10\% of all tweets---a singular, highly retweeted tweet
may contain 1-grams that are not produced organically by other users on that day
(e.g., a misspelling by a famous user).

We include some further figures as supplementary material.

We also provide a connected website associated with our paper at:\newline
\href{http://compstorylab.org/covid19ngrams/}{http://compstorylab.org/covid19ngrams/}.

We show tables of the leading 1-grams in our data set,
as well as an example bar chart race for the dominant COVID-19 1-grams in English.

Our intention is to automatically update the data set on Gitlab,
as soon as we have processed all tweets for a day.

\section{Concluding remarks}
\label{sec:covid19.concludingremarks}

We echo our main general observation of how COVID-19 has been discussed through
late March 2020:
\textbf{After reacting strongly in late January to the news that a coronavirus-based disease
was spreading in China, attention across all but 2 of the 24 languages we survey dropped
through February before resurging in late February and through March.}

We see abrupt shocks in time series as populations shifted rapidly to heightened
levels of awareness, particularly in the Italian time series.
In the time series for `virus', we see two and sometimes three peaks of attention
in the space of just a few months.

Our hope is that our collection of Twitter 1-gram time series
that are especially relevant to March 2020
will be of benefit to other researchers.

The time series we share will, in part, reflect many other aspects beyond
mentions of `virus', which we have only briefly explored here.
\textbf{Possible topics to investigate include
washing (including the soap emoji),
testing,
serology,
vaccine,
masks and protection equipment,
social and physical distancing,
terms of community support versus loneliness and isolation,
closures of schools and universities,
economic problems,
job loss,
and
food concerns.}

We repeat that the lists we provide are meant to represent
the important 1-grams of March 2020,
and we urge a degree of caution in the use of the data set.

As we have indicated above, our lists of 1-grams contain some peculiarities
that will not be directly relevant to COVID-19.
Entertainment (e.g., movies, celebrities, and K-pop) and sports (football along with sports in the United States)
are standard fare on Twitter when no major events are taking place in the world.
The extent to which these aspects of Twitter
are submerged as pandemic related 1-grams rise is of interest.

Finally, while we have been able to identify languages well,
geolocation is coarse and at best will be at the level of countries.
The strength of geolocation for our time series will
depend on the degree of localization of a given language
as well as Twitter user demographics.
We leave producing $n$-grams with serviceable physical location as
a separate project.

\acknowledgments
The authors are grateful for 
support furnished by MassMutual and Google,
and the computational facilities provided
by the Vermont Advanced Computing Core.
The authors appreciate discussions and correspondence with
Aaron Schwartz,
Todd DeLuca,
Nina Safavi,
and
Nicholas Danforth.

\clearpage

\renewcommand{\thefigure}{A\arabic{figure}}
\renewcommand{\thetable}{A\arabic{table}}
\setcounter{figure}{0}
\setcounter{table}{0}

\appendix

\section*{Appendix}

\begin{table}[!th]
  \centering
  \begin{tabular}{r|c}
    \textbf{Language} & \textbf{\# $1$-grams} \\
    \hline
        {English}    (en) &  3.75$\times 10^{9}$   \\
        {Spanish}    (es) &  1.41$\times 10^{9}$   \\
        {Portuguese} (pt) &  1.22$\times 10^{9}$   \\
        {Arabic}     (ar) &  8.34$\times 10^{8}$   \\
        {Korean}     (ko) &  1.10$\times 10^{9}$   \\
        {French}     (fr) &  2.18$\times 10^{8}$   \\
        {Indonesian} (id) &  3.32$\times 10^{8}$   \\
        {Turkish}    (tr) &  2.06$\times 10^{8}$   \\
        {German}     (de) &  1.17$\times 10^{8}$   \\
        {Italian}    (it) &  1.15$\times 10^{8}$   \\
        {Russian}    (ru) &  2.73$\times 10^{7}$   \\
        {Tagalog}    (tl) &  1.23$\times 10^{8}$   \\
        {Hindi}      (hi) &  8.05$\times 10^{8}$   \\
        {Persian}    (fa) &  1.22$\times 10^{8}$   \\
        {Urdu}       (ur) &  1.99$\times 10^{8}$   \\
        {Polish}     (pl) &  7.39$\times 10^{7}$   \\
        {Catalan}    (ca) &  1.37$\times 10^{8}$   \\
        {Dutch}      (nl) &  1.59$\times 10^{8}$   \\
        {Tamil}      (ta) &  1.30$\times 10^{8}$   \\
        {Greek}      (el) &  2.81$\times 10^{7}$   \\
        {Swedish}    (sv) &  1.91$\times 10^{7}$   \\
        {Serbian}    (sr) &  2.79$\times 10^{7}$   \\
        {Finnish}    (fi) &  1.34$\times 10^{7}$   \\
        {Ukrainian}  (uk) &  2.22$\times 10^{7}$   \\    
  \end{tabular}
  \caption{
    Approximate number of occurrences of all $1$-grams captured in our dataset by language
    for 2019/09/01 through to 2020/03/23.
    Languages are sorted by overall language popularity on Twitter for 2019.
  }
  \label{tab:covid19.1gramcounts}
\end{table}

\begin{figure*}
    \includegraphics[width=\textwidth]{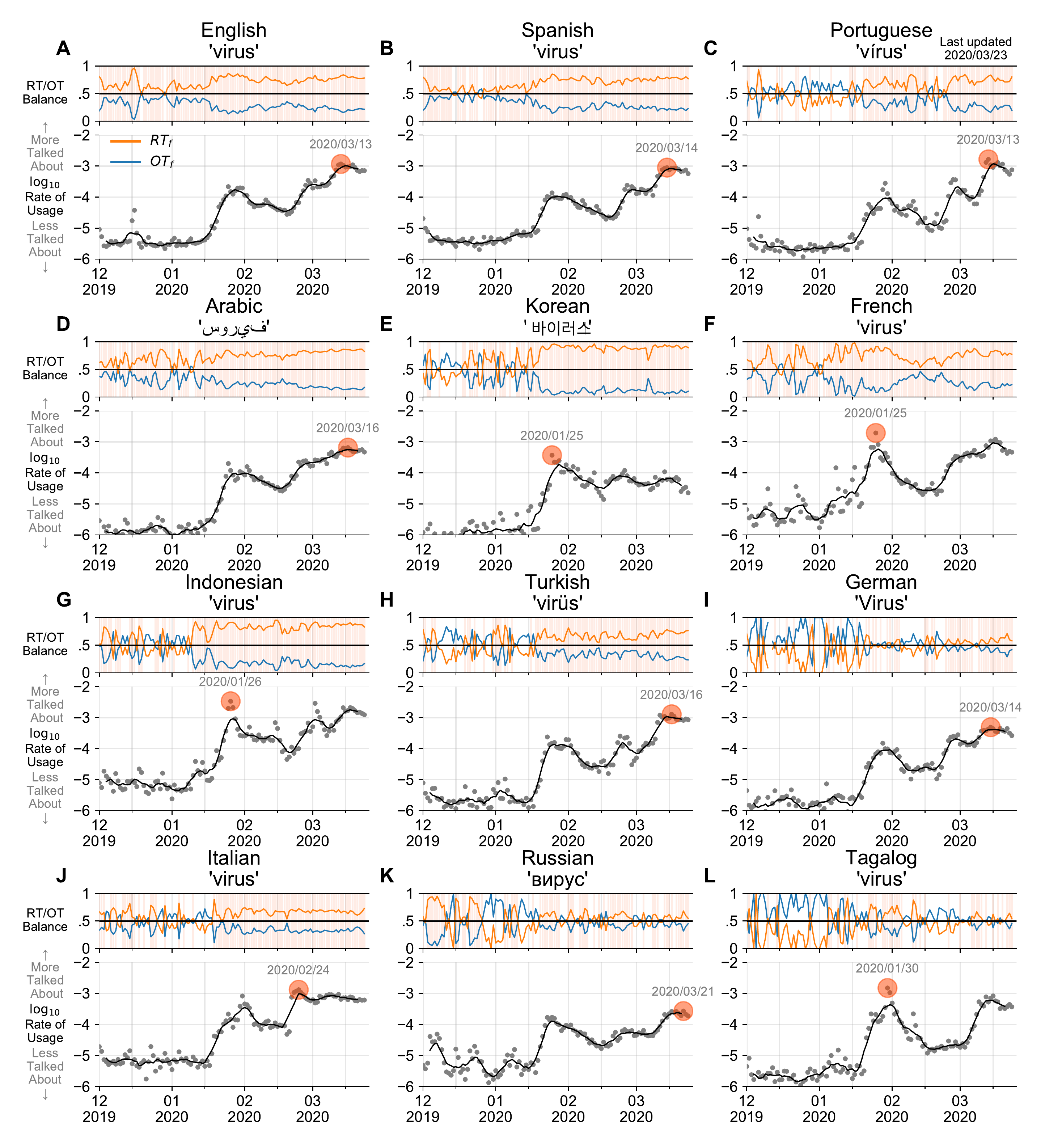}
    \caption{
      Relative usage rate time series for the word `virus' in the top 12 of the 24 languages we consider here,
      with language ranking by total usage in 2019.
      See the caption of Fig.~\ref{fig:covid19.virus1}
      as well as Sec.~\ref{sec:covid19.timeseries} for explanations of the plot formats.
      See the companion plots in Fig.~\ref{fig:covid19.virus1_ranks} for Zipfian rank time series.
    }
    \label{fig:covid19.virus1}
\end{figure*}

\begin{figure*}
    \includegraphics[width=\textwidth]{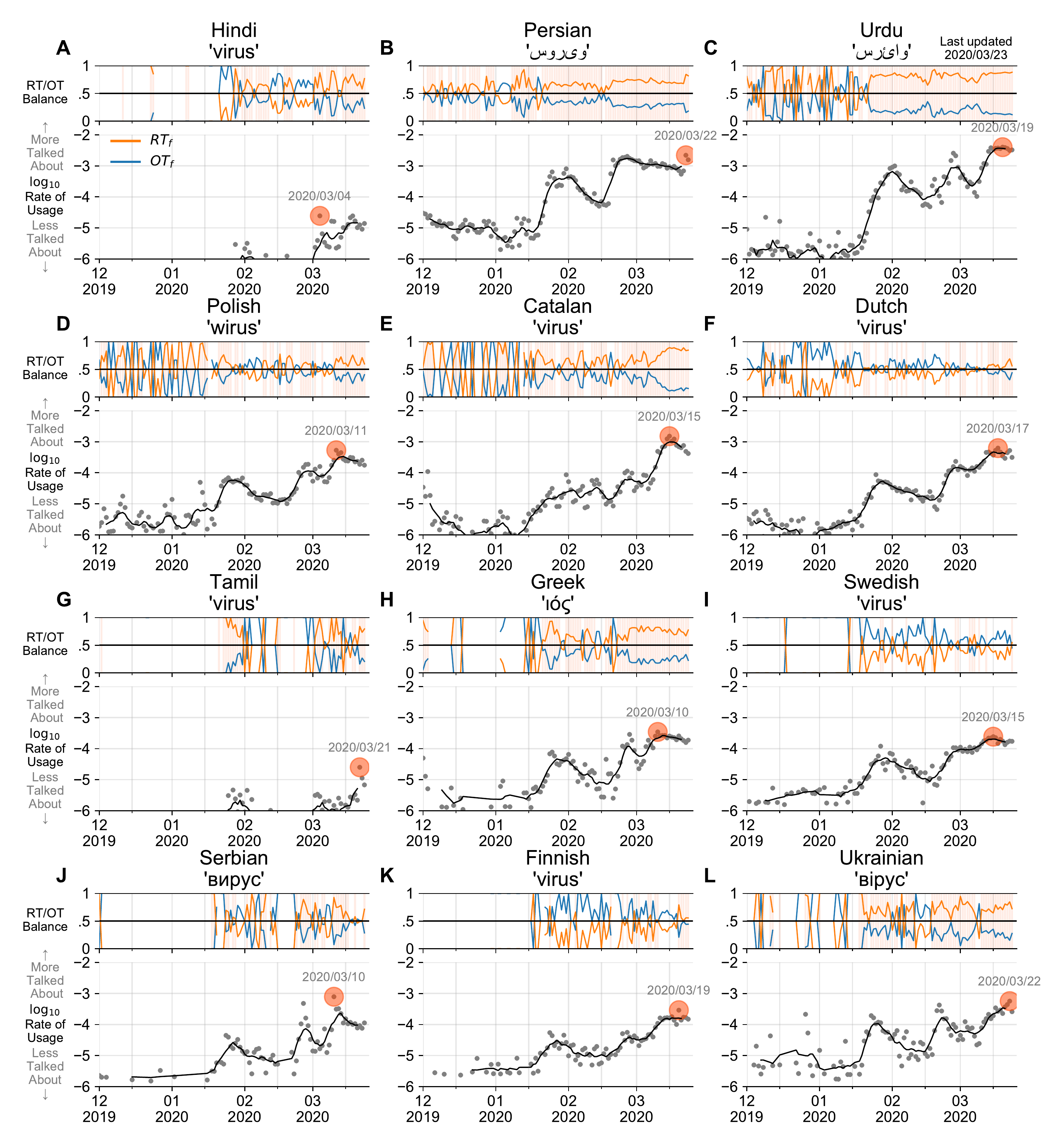}
    \caption{
      Continuing on from Fig.~\ref{fig:covid19.virus1}:
      Relative usage rate time series for the word `virus' in the bottom 12 of the 24 languages in our data set.
      See the companion plots in Fig.~\ref{fig:covid19.virus2_ranks} for Zipfian rank time series.
    }
    \label{fig:covid19.virus2}
\end{figure*}

\end{document}